\documentclass{emulateapj}

\usepackage[latin1]{inputenc}

\shorttitle{Balmer Lines and Complex BLR Properties}

\shortauthors{La Mura et al.}

\begin{abstract}
In this work we analyze a sample of AGN spectra, selected from the $6^{\rm
  th}$ Data Release of the Sloan Digital Sky Survey, exploiting a generalized
technique of line profile analysis, designed to take into account the whole
profiles of their broad emission lines. We find that the line profile
broadening functions result from a complex structure, but we may be able to
infer some constraints about the role of the geometrical factor, thus
improving our ability to estimate AGN properties and their relation with the
host galaxy. Our results suggest that flattening and inclination within the
structure of the Broad Line Region (BLR) must be taken into account. We detect
low inclinations of the BLR motion plane with respect to our line of sight,
typically $i \leq 20$°, with a geometrical effect which generally
decreases as the line profile becomes broader.
\end{abstract}

\newcommand{\ud}{\mathrm{d}}
\newcommand{\polstar}{$^\star$}

\begin{document}
\addtolength{\voffset}{0.5cm}
\title{Balmer Emission Line Profiles and the Complex Properties of
  Broad Line Regions in Active Galactic Nuclei}


\author{G. La Mura\altaffilmark{1}, F. Di Mille\altaffilmark{1},
  S. Ciroi\altaffilmark{1}, L. \v C. Popovi\'c\altaffilmark{2,3}, and
  P. Rafanelli\altaffilmark{1}}
\altaffiltext{1}{Department of Astronomy, University of Padova, Vicolo
  dell'Osservatorio, I-35122 Padova, Italy; giovanni.lamura@unipd.it,
  stefano.ciroi@unipd.it, francesco.dimille@unipd.it,
  piero.rafanelli@unipd.it.}
\altaffiltext{2}{Astronomical Observatory, Volgina 7, 11060 Belgrade,
  Serbia; lpopovic@aob.bg.ac.yu.}
\altaffiltext{3}{Isaac Newton Institute of Chile, Yugoslavia Branch,
  11060 Belgrade, Serbia.}

\maketitle

\section{Introduction}
During the past decades, spectroscopic observations provided a fundamental
starting point for our understanding of the physical processes in Active
Galactic Nuclei (AGN). In particular, the study of broad emission lines,
characterizing the spectra of many objects at different wavelengths, is a key
feature to penetrate their intrinsic properties \citep{Osterbrock89}.
Unfortunately, a correct interpretation of the observations requires
to collect information in various spectral ranges and for long
monitoring times, a task which is currently possible only in a fairly
small number of cases. While much work has been devoted to calibrate
empirical methods which should be able to deal with larger samples
\citep[see e. g.][]{Kaspi00, Kaspi05, Bentz06}, the fundamentally
unknown structure in the core of AGN influences the physical
interpretation of spectra with a geometrical factor, whose value
depends on the structure and orientation of the source \citep[see for
  instance][]{Vestergaard00, Nikolajuk05, Marziani06, Decarli08a}.

The prominent broad emission lines, visible in the spectra of many
AGN, originate close to the central power source, in the so called
Broad Line Region (BLR). Because of its small distance from the power
source, the BLR is in strong interaction with the radiation field
produced by the central engine and with its gravitational forces. Many
interesting details about the physics of processes that are taking
place within AGN can be identified in the signal of the BLR, but they
suffer from a still missing complete picture of the complex
kinematical and thermodynamical properties of the line emitting plasma. Since
it is not yet possible to directly observe the spatial distribution of the
broad line emitting medium, although many important achievements were obtained
in the angular resolution of AGN cores at radio wavelengths
\citep[e. g.][and references therein]{Kellerman98}, spectroscopic data are
still the most useful way to investigate physics within the BLR. The well
known Reverberation Mapping (RM) technique \citep[see][]{Blandford82},
based on multiple spectroscopic observations, provides a reliable way
to constrain the volume where the nuclear activity is confined and,
thus, to estimate the mass concentration therein \citep[e. g.][]
{Wandel99, Peterson00, Peterson04}.

The most commonly accepted interpretation of AGN puts a Super Massive
Black Hole (SMBH) in the role of the central engine, since matter
accretion into its gravitational field provides the required power to
account for many observational properties. Once the size of the BLR is
known, it is possible to estimate the mass of the SMBH:
$$M_{BH} = f\cdot\frac{R_{BLR} \Delta v^2}{G}, \eqno(1)$$
where $G$ is the gravitational constant, $f$ represents the geometrical
factor which accounts for the unknown distribution of the line emitting
material, $R_{BLR}$ is a characteristic BLR radius, and $\Delta v$ is
an estimate to the velocity field, usually coming from the width of
the emission lines. In their recent work, \citet{Kaspi05} found that a
power law relationship of the form $R_{BLR} \propto (5100\, \mbox{\AA}\,
L_{5100})^\alpha$ may adequately describe AGN, although the actual
value of the power law exponent and the luminosity range, where the
relationship holds in different AGN classes, still have to be
constrained \citep{Kaspi07, Kelly07, Laor07, McGill08}.
\citet{Bentz06} argued that, at least for moderate luminosity sources,
with $L_{bol} \leq 10^{46}\, {\rm erg\, s^{-1}}$, it is likely that
$\alpha \simeq 0.52$, not far from the predictions of simple
photoionization calculations giving $\alpha \sim 0.5$.

Unfortunately, the problems introduced by our limited knowledge about
the actual structure of the BLR badly affect the value of such
estimates, rising many uncertainties which make it quite difficult to
infer the physical properties of AGN or to study their relation with
the host galaxy environment. During the past years, a lot of work has
been devoted to understand the relationship between the BLR dynamics
and the corresponding broad emission line profiles
\citep[e. g.][]{Capriotti80,   Capriotti81, Ferland92, Peterson99,
  Korista04}, but, while the former is probably very complex, often
with evidence for multiple  components \citep{Popovic04}, the latter
is the result of a combination of effects involving the gas motion
pattern and the radiation transfer across an environment which is only
approximately understood.

In this paper we describe the results we obtained by analyzing the
kinematical properties of the BLR gas in a sample of AGN extracted
from the Sloan Digital Sky Survey (SDSS) database, by means of a
technique exploiting the cross-correlation method and the
Gauss-Hermite profile fitting to infer the line broadening function
(BF) in the optical domain. We show that some interesting clues to the
geometry of the BLR can be identified in this way and we apply the
results to estimate the physical properties in our sample of AGN.

The paper is organized as follows: in \S2 we describe the analytical
formalism to extract the profile broadening functions and to calculate
the corresponding Gauss-Hermite expansions; in \S3 we present our
sample and the reduction techniques that we adopted; \S4 summarizes
our results, with a discussion of the main limits and some indications
to improve the analysis; finally our conclusions are given in \S5.

\section{Line profile analysis and theoretical models}
In the effort towards revealing the intrinsic properties of AGN cores,
line profile analysis often played a major role. Under specific
assumptions about the dynamical conditions within the BLR, such as the
hypothesis of virial motions driven by the combined effect of the
central engine's gravity and radiation pressure, a number of
representative parameters, like the line widths at different intensity
levels, or the line asymmetry factors, usually computed in the form of
ratios among the line extension toward the blue and red wavelengths,
with respect to the line core position, were used in order to describe
the profiles and to evaluate the properties of the engine. This kind
of approach is prone to the effects of the substantially unknown BLR
geometry, with the possibility to introduce systematic misinterpretation of
data. Furthermore, it assumes quite specific measurements to be a good
approximation of the entire emission line profile, loosing some
precious physical details. 

In this section we describe a generalized approach to the line profile
fitting, already exploited in the past years in the field of advanced
stellar kinematics, but adopted for gas kinematics as well
\citep{Barton00}.

\begin{deluxetable*}{lcccccc}
\tabletypesize{\footnotesize}
\tablewidth{0pt}
\tablecaption{List of objects in the sample \label{tab01}}
\tablehead{\colhead{Name} & \colhead{RA} & \colhead{Dec} &
  \colhead{M$_{\rm V}$} & \colhead{z} & \colhead{FWHM$_{\rm H\beta}$}
  & \colhead{Detected}\\ \colhead{} & \colhead{(hh:mm:ss.s)}
  & \colhead{(dd:mm:ss)} & \colhead{} & \colhead{} & \colhead{(${\rm
      km\, s}^{-1}$)} & \colhead{in Radio}}
\startdata
RX J0801.5+4736           & 08:01:32.0 & +47:36:16 & -23.3 & 0.157 & 7009 $\pm$ 336 & Yes \\
RXS J080358.9+433248      & 08:03:59.3 & +43:32:58 & -24.4 & 0.451 & 2632 $\pm$ 222 & No \\
1RXS J080534.6+543132     & 08:05:34.9 & +54:31:29 & -26.5 & 0.406 & 3174 $\pm$ 299 & No \\
SDSS J081222.99+461529.1  & 08:12:23.0 & +46:15:28 & -23.6 & 0.311 & 2807 $\pm$ 283 & No \\
2MASSi J0816522+425829    & 08:16:52.3 & +42:58:30 & -23.1 & 0.235 & 3768 $\pm$ 186 & No \\
NGC 2639 U10              & 08:42:30.5 & +49:58:03 & -23.3 & 0.305 & 6046 $\pm$ 210 & No \\
SDSS J085632.39+504114.0  & 08:56:32.4 & +50:41:14 & -23.9 & 0.234 & 2819 $\pm$ 157 & No \\
SDSS J085828.69+342343.8  & 08:58:28.6 & +34:23:44 & -24.3 & 0.257 & 3570 $\pm$ 265 & No \\
SDSS J090455.00+511444.6  & 09:04:55.0 & +51:14:44 & -23.4 & 0.224 & 3620 $\pm$ 142 & No \\
RX J0906.0+4851           & 09:06:01.3 & +48:51:48 & -24.1 & 0.390 & 2503 $\pm$ 100 & No \\
RX J0908.7+4939           & 09:08:47.4 & +49:40:07 & -24.5 & 0.421 & 2137 $\pm$  71 & No \\
1WGA J0931.9+5533         & 09:32:00.1 & +55:33:48 & -23.1 & 0.265 & 5045 $\pm$ 482 & Yes \\
SDSS J093653.84+533126.8  & 09:36:53.8 & +53:31:27 & -23.1 & 0.228 & 4255 $\pm$ 175 & No \\
FIRST J094610.9+322325    & 09:46:10.9 & +32:23:26 & -24.5 & 0.403 & 1684 $\pm$ 235 & Yes \\
KUV 09484+3557            & 09:51:23.8 & +35:42:48 & -23.8 & 0.398 & 3244 $\pm$ 289 & No \\
HS 1001+4840              & 10:04:13.8 & +48:26:06 & -24.0 & 0.562 & 2962 $\pm$ 230 & No \\
PC 1014+4717              & 10:17:30.9 & +47:02:25 & -23.7 & 0.335 & 2583 $\pm$ 103 & No \\
RX J1030.4+5516           & 10:30:24.9 & +55:16:21 & -24.8 & 0.435 & 2246 $\pm$ 157 & Yes \\
FBQS J103359.4+355509     & 10:33:59.5 & +35:55:09 & -23.3 & 0.169 & 4543 $\pm$ 237 & Yes \\
SBS 1047+557B             & 10:50:55.1 & +55:27:23 & -24.1 & 0.333 & 2253 $\pm$  97 & No \\
RX J1054.7+4831           & 10:54:44.7 & +48:31:39 & -24.7 & 0.286 & 4428 $\pm$ 274 & Yes \\
FBQS J105648.1+370450     & 10:56:48.2 & +37:04:51 & -24.1 & 0.387 & 4538 $\pm$ 272 & Yes \\
FBQS J110704.5+320630     & 11:07:04.5 & +32:06:30 & -23.3 & 0.243 & 8447 $\pm$ 558 & Yes \\
FBQS J112956.5+364919     & 11:29:56.5 & +36:49:19 & -24.4 & 0.399 & 3564 $\pm$ 452 & Yes \\
FBQS J115117.7+382221     & 11:51:17.8 & +38:22:21 & -25.4 & 0.336 & 3696 $\pm$ 242 & Yes \\
RX J1200.4+3334           & 12:00:28.7 & +33:34:43 & -23.5 & 0.284 & 2864 $\pm$ 173 & No \\
RX J1203.8+3711           & 12:03:54.8 & +37:11:37 & -23.4 & 0.401 & 3725 $\pm$ 1020 & Yes \\
1RXS J121759.9+303306     & 12:17:59.3 & +30:33:03 & -23.8 & 0.363 & 3435 $\pm$ 258 & No \\
RX J1218.3+3850           & 12:18:22.6 & +38:50:43 & -24.4 & 0.194 & 4564 $\pm$ 483 & No \\
FBQS J122035.1+385317     & 12:20:35.1 & +38:53:17 & -25.3 & 0.376 & 1899 $\pm$ 164 & Yes \\
FBQS J122424.2+401510     & 12:24:24.2 & +40:15:10 & -24.3 & 0.415 & 4721 $\pm$ 397 & Yes \\
FBQS J122624.2+324429     & 12:26:24.2 & +32:44:29 & -23.9 & 0.242 & 4535 $\pm$ 533 & Yes \\
FBQS J125602.0+385230     & 12:56:02.0 & +38:52:30 & -24.1 & 0.419 & 1978 $\pm$ 147 & Yes \\
FBQS J132515.0+330556     & 13:25:15.0 & +33:05:56 & -23.5 & 0.356 & 1921 $\pm$ 102 & Yes \\
SDSS J144050.77+520446.0  & 14:40:50.7 & +52:04:46 & -23.4 & 0.320 & 2483 $\pm$ 134 & No \\
RX J1452.4+4522           & 14:52:24.7 & +45:22:24 & -24.6 & 0.469 & 6594 $\pm$ 156 & Yes \\
FBQS J145958.4+333701     & 14:59:58.4 & +33:37:01 & -25.5 & 0.644 & 4389 $\pm$ 209 & Yes \\
FIRST J154348.6+401324    & 15:43:48.7 & +40:13:25 & -23.4 & 0.318 & 5227 $\pm$ 391 & Yes \\
SDSS J154833.03+442226.0  & 15:48:33.1 & +44:22:26 & -23.5 & 0.322 & 3893 $\pm$ 343 & Yes \\
FBQS J155147.4+330007     & 15:51:47.4 & +33:00:08 & -24.5 & 0.422 & 2744 $\pm$ 221 & Yes \\
\enddata
\tablecomments{Together with names and equatorial coordinates (J2000.0),
  the table lists the absolute magnitude in the {\it V} band, the
  cosmological redshift, FWHM$_{\rm H\beta}$, and a binary flag
  indicating whether the source is detected in radio or not.}
\end{deluxetable*}

\subsection{Emission line broadening from cross-correlation}
The BLR spectrum shows several emission lines corresponding to many
permitted and some semi-forbidden transitions of variously ionized
atomic species. There are indications that the distribution of the
line emitting material is different, according to the ionization
potential of the considered emission lines \citep[see
  e. g.][etc.]{Gaskell86, Sulentic95, Marziani96, Snedden04,
  Matsuoka08, Mullaney08, Sluse08}. Indeed, the interaction of the 
most energetic AGN radiation with gas probably produces a region where
matter is highly ionized. On the contrary, optical shielding effects
allow for the survival of low ionization species, in regions where
only comparatively low energy photons may penetrate. Therefore, if the
BLR structure is such that the shielded component is different from
the directly exposed one, the properties of the emission lines will
depend on their ionization potential.

On the other hand, choosing to analyze a set of emission lines belonging to a
statistical distribution of matter and radiation interactions, it is more
likely that the emission regions are not dramatically different. In the
optical domain, the Balmer series of hydrogen is the most appropriate choice,
because of its strength above the underlying continuum.

Assuming that the Balmer line emission is not affected by large
variations across the BLR and introducing the cross-correlation
formalism, originally described by \citet{Tonry79} and then updated by
\citet{Statler95}, we can approximate the observed line spectra as the
convolution of an appropriate template of narrow emission lines $T(x)$ with
the BLR broadening function $B(x)$:
$$S(x) \simeq T(x) \ast B(x), \eqno(2)$$
where $S(x)$ is the observed spectrum, while $x$ represents a logarithmic
wavelength coordinate of the form:
$$x = A \ln\lambda + B, \eqno(3)$$
such that the effect of radial velocities results in linear shifts
along $x$. Eq.~(2) can be explicitly written as:
$$S(x) \simeq \int T(x) B(x - x')\, \ud x' \eqno(4)$$
and, if we compute the cross-correlation function of the spectrum with
the template, we find:
$$X(x) = S(x) \otimes T(x) = \int S(x) T(x + x')\, \ud x'. \eqno(5)$$
Using Eq.~(4), the cross-correlation function becomes:
$$X(x) \simeq \int\,\int T(x) B(x - x'')\, \ud x'' T(x + x')\, \ud x',
\eqno(6)$$
which, upon changing order of integration, is:
$$X(x) \simeq \int\,\int T(x) T(x + x')\, \ud x' B(x - x'')\, \ud x''.
\eqno(7)$$
Based on the definitions of cross-correlation and convolution, Eq.~(7)
approximates the cross-correlation function of the spectrum and the
template as the convolution of the template autocorrelation function
with the object's BF \citep{Statler95}:
$$X(x) \simeq [T(x) \otimes T(x)] \ast B(x). \eqno(8)$$
Since $T(x)$ is known and $X(x)$ is drawn from observations, as far as
the template is correct, it is possible to recover $B(x)$. 

Restricting our analysis to the primary cross-correlation peak, which
carries most of the kinematical information and it is weakly affected
by template mismatch, Eq.~(7) can be written in its discrete
form, with the simplified notation $F_i = F(x_i)$:
$$X_k \simeq \sum_{i = 0}^N \left(\sum_{j = 0}^N T_i\, T_{i + j}\right)
B_{k - i}. \eqno (9)$$
Provided that all the functions are null when they are computed
outside the range $0 \leq i \leq N$, Eq.~(9) defines a system of
$N + 1$ linear equations in the $N + 1$ variables $B_{k - i}$ ($k \geq
i$). A standard $\chi^2$-minimization routine can be therefore used to
infer the BF of the Balmer lines.

\subsection{Analytical expressions for the broadening functions}
As previously mentioned, the BLR broadening functions are influenced
by the effects of complex kinematics within the source and of
radiation transfer from the source to the observer. For this reason it
is hardly conceivable that a simple analytic expression might be used
to fit the resulting profiles. In the case of a geometrically complex
distribution of motions in the line emitting region, multiple Gaussian
functions provide reasonable fits to the observed profiles. Two
Gaussian contributions can usually fit the broad component of
H$\beta$ \citep{Popovic04, Chen08}, but other contributions, up to
five more Gaussians, might be needed to account for the narrow
emission lines of H$\beta$ and [\ion{O}{3}]. Furthermore, the presence
of ordered kinematical components modifies the shape of the BF,
raising non-Gaussian features in the profiles.

A good way to estimate the importance of non-Gaussian components is to
parameterize the observed BF by means of a Gauss-Hermite orthonormal
expansion, similarly to what is described in \citet{VanDerMarel93} for
the case of stellar kinematics in elliptical galaxies. Following their
method, if we call $\alpha(v)$ the normal Gaussian function:
$$\alpha(v) = \frac{1}{\sqrt{2\pi}\sigma_v}\exp\left(-\frac{v^2}{2\sigma_v^2}
\right), \eqno(10)$$
where $\sigma_v$ is the line of sight velocity dispersion, the
emission line BF can be expressed as a function of {\it v\ }:
$$B(v) = B_0 \alpha(v - V_{sys}) \left[1 + \sum_{i = 3}^N h_i
  H_i(v - V_{sys})\right], \eqno(11)$$
in which we call $B_0$ the BF normalization factor, $V_{sys}$ the
systemic radial velocity offset between the BF and the chosen 
reference frame, $H_i(v - V_{sys})$ the $i^{\rm th}$ order Hermite
polynomial, and $h_i$ the corresponding coefficient. A wide
description of the properties of the Hermite polynomials is given in
\citet{VanDerMarel93}. It is demonstrated that odd order functions account for
asymmetric distortions of the Gaussian profile, while even order functions
have a symmetric effect. Truncating Eq.~(11) to $N = 4$, the Hermite
polynomials are expressed by:
$$H_3(y) = \frac{1}{\sqrt{6}}(2\sqrt{2} y^3 - 3\sqrt{2} y) \eqno(12a)$$
$$H_4(y) = \frac{1}{\sqrt{24}}(4 y^4 - 12 y^2 + 3). \eqno(12b)$$
Therefore, it is possible to estimate the role of non-Gaussian kinematical
components, using the whole BF profile, simply by fitting the observed shape
with a truncated Gauss-Hermite series and measuring the appropriate values of
$h_3$ and $h_4$.

\section{Sample selection and data reduction}
\subsection{AGN spectra from SDSS}
To perform our investigation, we needed a homogeneous sample of AGN optical
spectra, featuring prominent broad Balmer emission lines. The spectroscopic
database of the $6^{th}$ data release of SDSS (DR6) provides a huge number of
objects, whose spectra are collected and processed by a fairly well
established standard pipeline \citep{SDSSDR6}. In order to collect a
set of good signal spectra, we chose to select the sample in the
V\'eron-Cetty catalogue of Quasars and Active Galactic Nuclei
(12$^{th}$ Ed.) \citep{VeronCetty06}, with the following
requirements:
\begin{itemize}
\item the object redshift had not to exceed the limit $z \approx 0.8$,
  because objects at larger redshift have their H$\beta$ emission line beyond
  the SDSS spectral coverage;
\item only bright sources, with ${\rm M_V} < -23$, were considered;
\item each object had a spectral classification suggesting the presence of
  broad emission line profiles;
\item the sources are located within the field covered by the SDSS DR6
  spectroscopic observations.
\end{itemize}

On the resulting candidate list, we applied further constraints, that
restricted the sample to the most appropriate sources. In particular we chose
objects whose spectra had at least three clearly detectable broad Balmer lines
and they were not affected by instrumental disturbances or by strong
foreground and background contamination. We ended up with a sample of 40
objects that is described in Table~1.

\begin{figure}[t]
\includegraphics[width = 8.3cm]{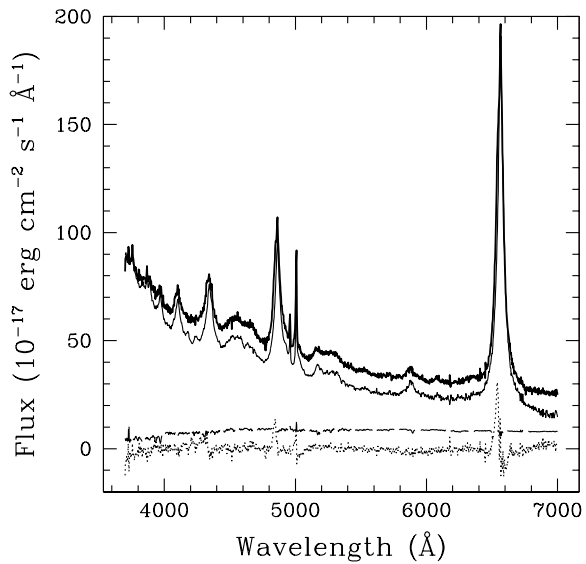}
\caption{Example of spectral decomposition for SDSS
  J085632.39+504114.0. The observed spectrum, represented by the thick
  continuous line, is compared with the AGN (thin continuous line) and
  galactic (long dashed) components. The dotted line in the bottom part of the
  plot is the fit residual. \label{f01}}
\end{figure}
\subsection{Preliminary spectral reduction}
A major advantage in the SDSS database is that it provides spectra
with preliminary reduction and calibration, thus simplifying the task
of spectral analysis. Therefore, before proceeding with our
measurements, we simply had to remove from the spectra those
contributions which come from outside the AGN. We applied a
correction for Galactic Extinction, estimated by means of a selective
extinction function, in the form proposed by \citet{Cardelli89} with the
absorption coefficients evaluated on the basis of the extinction map
traced by \citet{Schlegel98} and available at the NASA Extragalactic
Extinction Calculator. We, then, removed the cosmological redshift,
bringing the spectra to the rest frame of the narrow [\ion{O}{3}] emission
lines. Finally, we estimated the host galaxy contamination by applying
a spectral decomposition technique, based on the Karhunen-Lo\'eve Transforms
described by \citet{Connolly95} and implemented on SDSS data by
\citet{VandenBerk06}. According to the same method exploited in
\citet{LaMura07}, we consider the observed spectra as the linear combinations
of principal components, called {\it eigenspectra}, originated independently
by the AGN and its host:
$$S(\lambda) = \sum_{i = 1}^n[q_i\cdot Q_i(\lambda)] + \sum_{j = 1}^m[g_j\cdot
G_j(\lambda)], \eqno(13)$$
with $S(\lambda)$ being the total spectrum, $Q_i(\lambda)$ the $i^{\rm th}$
AGN component, weighted by its coefficient $q_i$, and $G_j(\lambda)$ the
$j^{\rm th}$ host galaxy eigenspectrum associated to the corresponding
coefficient $g_j$. Using the galactic and AGN eigenspectra provided by
\citet{Yip04a, Yip04b}, we evaluated the coefficients ($q_i$, $g_j$) by
means of a $\chi^2$-minimization routine involving the first five galaxy
eigenspectra and six AGN components. This procedure allows to carry out the
separation outlined in Eq.~(13), as it is illustrated in Fig.~1, and to
subtract an estimate of the host galaxy starlight contaminating the SDSS
spectra.

\begin{figure*}[t]
\begin{center}
\includegraphics[width = 12.7cm]{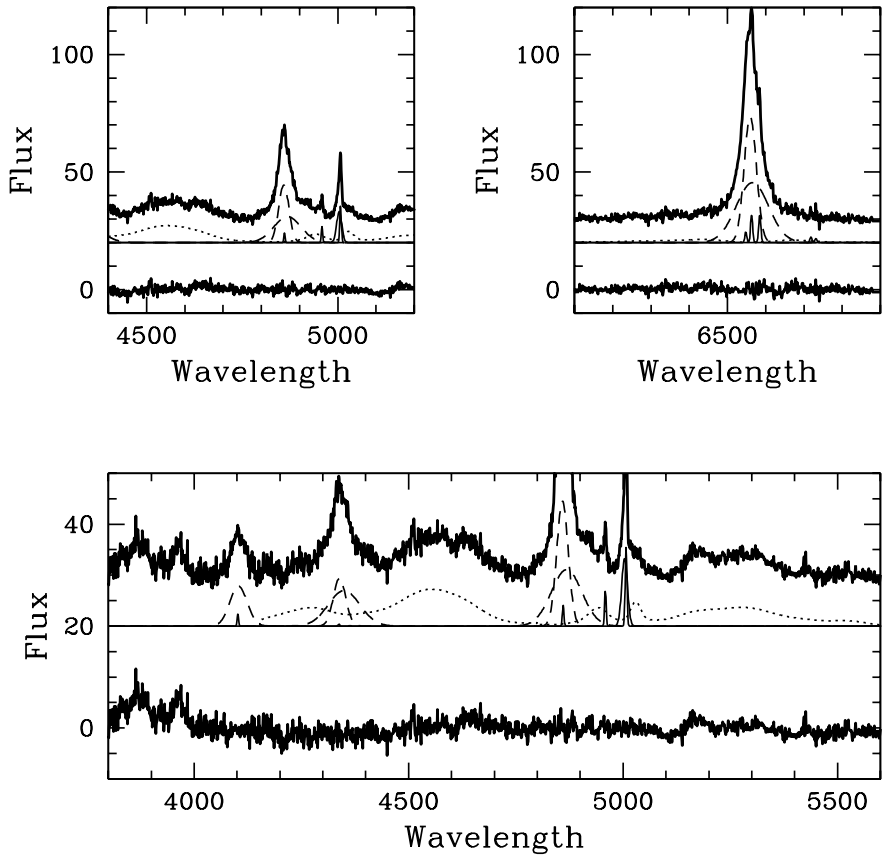}
\end{center}
\caption{Multiple Gaussian decompositions of the profiles of H$\beta$
  (upper left panel), H$\alpha$ (upper right), H$\gamma$, and
  H$\delta$ (lower panel). Here we use a thick continuous line to plot
  the continuum subtracted spectrum of PC~1014+4717, a thin continuous
  line for the estimated NLR contributions, a long dashed line for the
  BLR components, and a dotted line for \ion{Fe}{2}. The thick
  continuous line in the bottom part of each panel shows the fit
  residuals, while fluxes and wavelengths are expressed in units of
  $10^{-17} \mbox{erg}\, \mbox{cm}^{-2}\, \mbox{s}^{-1}\,
  \mbox{\AA}^{-1}$ and \AA, respectively. \label{f02}}
\end{figure*}
\subsection{Extracting the BLR component}
Once the AGN spectra have been corrected to account for most of the external
effects, the task to identify the BLR contribution alone needs the removal of
more components, including the underlying continuum of the AGN central source
and the narrow lines coming from the Narrow Line Region (NLR). Measuring the
properties of the broad lines, moreover, is often difficult because of the
multiple spectral features that are blended together. In the case of the
Balmer series, narrow lines from [\ion{O}{3}], [\ion{N}{2}], [\ion{S}{2}],
together with the narrow Balmer emissions, have to be taken into account,
while blends with broad lines from \ion{He}{2} and the multiplets of
\ion{Fe}{2} can heavily affect the observed profiles.

To subtract the AGN continuum, we fit the spectra with {\tt spline} functions
of order ranging from 2 to 5 in wavelength ranges which are usually not
affected by prominent lines. The subtraction of narrow lines and the blends
with He are dealt with by means of multiple Gaussian profile decompositions,
where we use the [\ion{O}{3}] emission line at 5007 \AA\ as a template
for the other narrow features and we fix the [\ion{N}{2}]
$\lambda\lambda$ 6548, 6584 emission line ratio to be 1:3. Some
special care, instead, is needed in the case of the \ion{Fe}{2}
multiplets, whose properties have not yet been understood in
detail. Following the method suggested by \citet{VeronCetty04}, it is
possible to remove the \ion{Fe}{2} contribution from spectra by
scaling and broadening an appropriate template, estimated from the
spectrum of an AGN featuring prominent \ion{Fe}{2} emission lines. In
our work, we used the \ion{Fe}{2} template spectrum coming from {\it I
  Zwicky} 1 \citep{Botte04}, splitting it into two parts, which we
scaled in order to achieve a better coincidence with our data. The
process is summarized in Fig.~2. Most of the steps so far described
were carried out with tasks provided in the {\it IRAF} software
package.

Once the removal of contaminating contributions has been performed, we
are able to apply the Gauss-Hermite formalism to the profiles of the
broad lines so far isolated in the spectra. We perform a first
analysis directly on the profiles of the H$\beta$ emission line. The
results of this study can be subsequently compared with the shape of
the Balmer line broadening functions, which we are now able to infer
from the cleaned BLR spectra. 

\begin{figure*}[t]
\begin{center}
\includegraphics[width = 6.3cm]{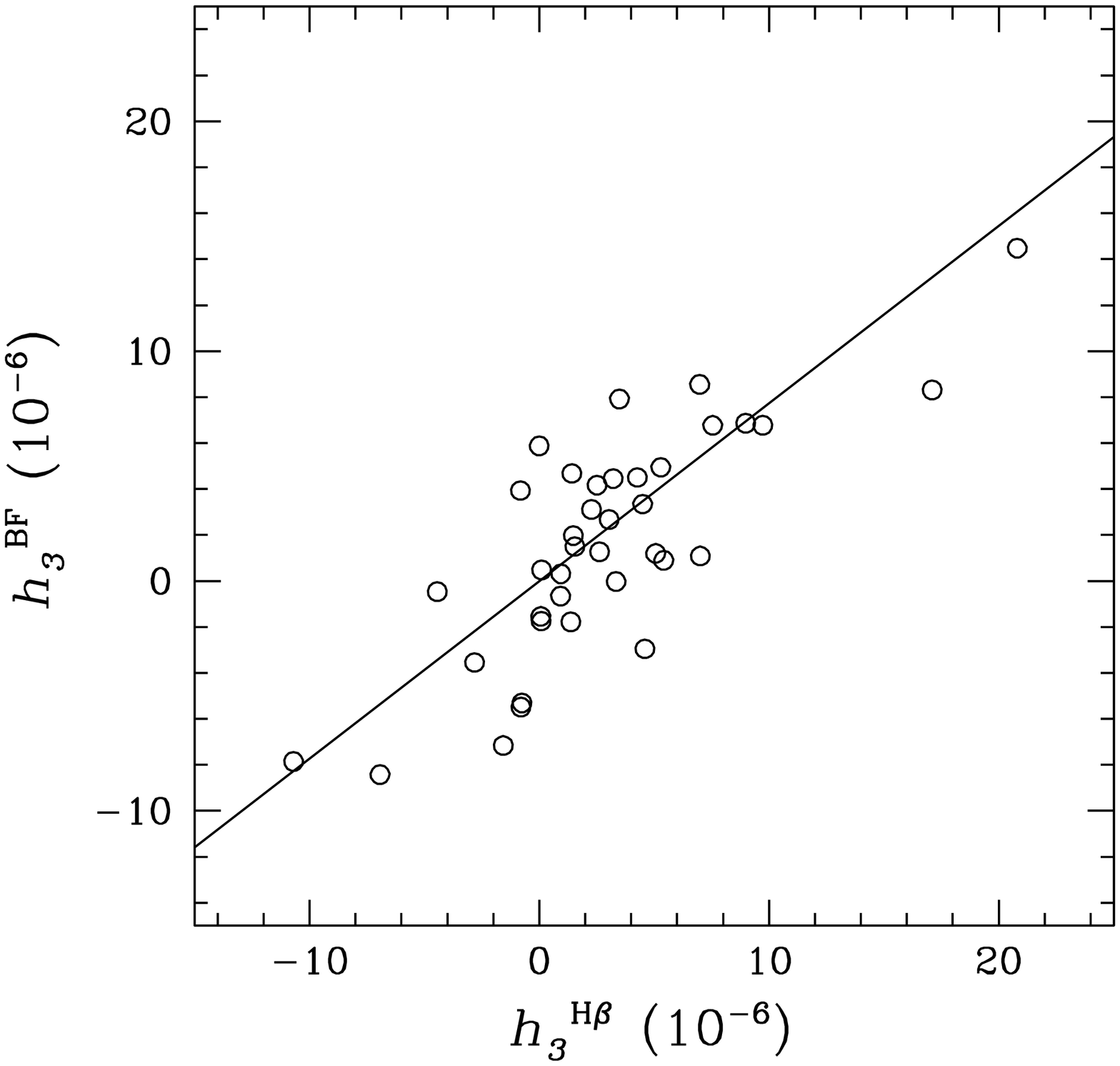}
\includegraphics[width = 6.3cm]{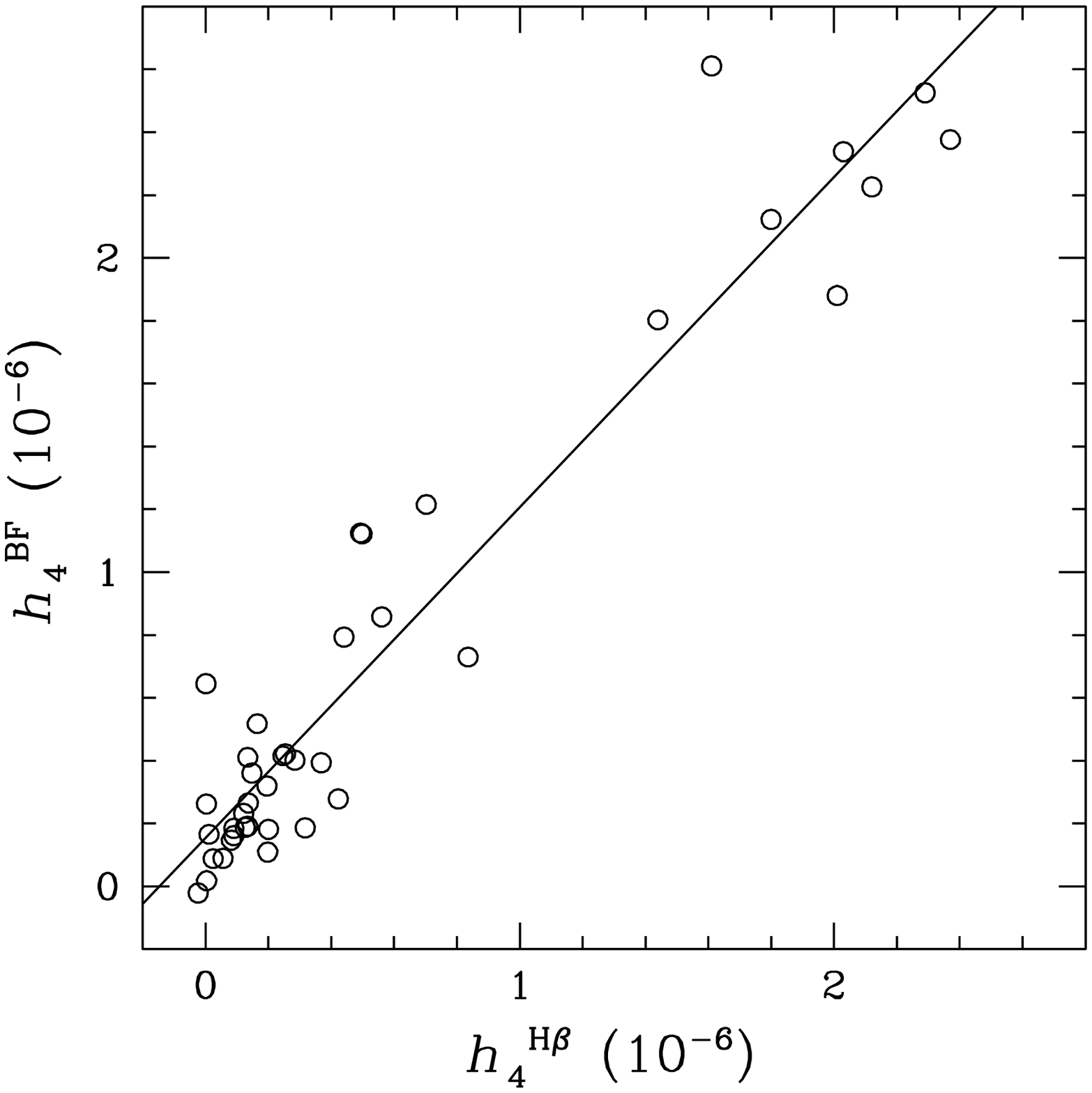}
\end{center}
\caption{Gauss-Hermite expansion coefficients of the line profile BF compared
  with the same coefficients extracted from the profile of H$\beta$. The
  straight lines shown in both panels are the best fit functions described in
  Eq.~(16) and Eq.~(17). \label{f03}}
\end{figure*}
\subsection{The Balmer Line Broadening Functions}
As a result of the previous steps, we now have got a set of BLR Balmer
line spectra. Our task is then to recover their BF, by  means of the
cross-correlation technique outlined in \S2.1. To calculate the
cross-correlation functions, we build a template of Balmer emission
lines, following the median line intensity ratios found by
\citet{LaMura07} in a sample of 90 SDSS spectra of broad line emitting
AGN. Our template assumes that the SDSS instrumental profile is a
Gaussian function with FWHM $= 167\, {\rm km\, s^{-1}}$. At the
spectral resolution of Sloan data, the logarithmic sampling of the
wavelength coordinate can be performed with discrete bins
corresponding to 69 ${\rm km\, s^{-1}}$ each. Here we use the {\it
  IRAF} task {\tt   fxcor} to compute the template autocorrelation
function:
$$A(x) = T(x) \otimes T(x) \eqno(14)$$
and the cross-correlation functions of the BLR spectra with the template,
following the definition of Eq.~(5). Again a $\chi^2$-minimization
algorithm can be exploited to infer the BF in its discrete form. Applying the
least squares formalism to the equation system~(9), it follows that the BF of
each spectrum must satisfy the relations:
$$\sum_{i = 0}^N B_i \left(\sum_{j = 0}^N A_j A_{i - k}\right) = \sum_{i =
  k}^N A_{i - k} X_i. \eqno(15)$$

In principle, it is possible to extract an accurate solution for the BF by
solving the equation system~(15) with $0 \leq k \leq N$. In practice the task
is not simple, because it involves the inversion of a coefficient matrix as
large as $[(N + 1)\times (N + 1)]$, with $N$ increasing with the line profile
widths up to $N \simeq 400$. However, the complete solution of such a
system is not the real purpose of this work, since we are not seeking the
detailed shape of the BF, but we are rather looking for the importance of
non-Gaussian components. Therefore, we chose to solve the system at lower
resolution, interpolating the BF every 8 bins with an analytical profile,
which we assumed to be a Gauss-Hermite expansion. 

We compared the properties inferred for the BF of our spectra with the results
obtained by applying the Gauss-Hermite profile fitting directly to the
H$\beta$ emission line. We found that the expansion coefficients in the two
cases are highly correlated, supporting a tight relationship among the
H$\beta$ emission line profile and the BF of the Balmer series. The
details of this comparison are given in Fig.~3, where we plot the
values of the expansion coefficients obtained in both ways. As a
result, we get:
$$h_3^{\rm BF} = (0.773 \pm 0.073) h_3^{\rm H\beta} + (0.721 \pm 0.554) \cdot
10^{-6}, \eqno(16)$$
with a correlation coefficient $R = 0.865$ and a null hypothesis $P_0 <
10^{-6}$ and
$$h_4^{\rm BF} = (1.051 \pm 0.052) h_4^{\rm H\beta} + (0.154 \pm 0.049) \cdot
10^{-6}, \eqno(17)$$
with $R = 0.955$ and $P_0 < 10^{-6}$.

Here we would like to point out that the matrix inversion must be
computed only once, because it involves coefficients exclusively drawn
from the template autocorrelation function. The cross-correlation
functions of the spectra, instead, only affect the known terms 
of Eq.~(15). Hence the advantage of this technique.

\subsection{Spectral property measurements}
After the calculation of the BF of our sample, we performed more
measurements of spectral properties in the data, estimating, in
particular, the FWHM of the H$\beta$ emission line and the AGN
continuum radiation luminosity at 5100 \AA. These are needed to infer
some of the source physical properties, such as the central black hole
mass, its accretion rate and the size of the BLR.

In the case of the line profile measurements, we looked at the
H$\beta$ emission line in the BLR spectra, which we previously
isolated for cross-correlation with the template. To account 
for the uncertain continuum and narrow line corrections, we performed five
different measurements of the line half width at half the maximum both on the
blue and red wing of the line, varying our guess to the continuum and line
peak intensity. We combined these estimates to calculate the FWHM, then we
averaged them together, and we computed the $1\sigma$ dispersion. As a further
step, we fit the broad H$\beta$ profile with two Gaussian functions and
we applied similar measurements to the identified components.

The optical continuum luminosities at 5100 \AA, instead, were evaluated in
the AGN spectra that we previously corrected for Galactic Extinction and host
contamination. Here, the main source of error arises from the noise
fluctuations around the actual signal intensity. For this reason, we assumed
the specific continuum luminosity at 5100 \AA\ to be represented by
the average luminosity, evaluated in the range running from 5075 \AA\ to
5125 \AA, and the associated error to be given by its standard
deviation. Therefore, we measured the continuum fluxes of our spectra and we
computed the related specific luminosities using the object redshift as a
distance indicator, in the framework of a cosmological model defined by $H_0 =
75\, {\rm km\, s^{-1}\, Mpc^{-1}}$, $\Omega_{matter} = 0.3$, and
$\Omega_\Lambda = 0.7$. A guess to the bolometric luminosity, which is
needed to infer the accretion rate onto the central black hole, can be
made from the specific luminosity measured in the spectra
\citep[e. g.][and references therein]{Elvis94, Collin02bsa,
  Collin04}. Here we assume our objects to have bolometric
luminosities approximately given by:
$$L_{bol} \simeq 10\cdot (5100\, \mbox{\AA}\cdot L_{5100}), \eqno(18)$$
where $L_{5100}$ represents our estimate to the continuum specific
luminosity at 5100 \AA, measured in ${\rm erg\, s^{-1}}$ \AA$^{-1}$,
and $L_{bol}$ is the bolometric luminosity, given in ${\rm erg\, s^{-1}}$.
We note, however, that this assumption is prone to the effect of the
large dispersion in the SED of AGN and it may easily introduce an
uncertainty of a factor $\sim 2$, which propagates in the estimated
black hole masses and accretion rates.

\begin{figure}[t]
\includegraphics[width = 8.3cm]{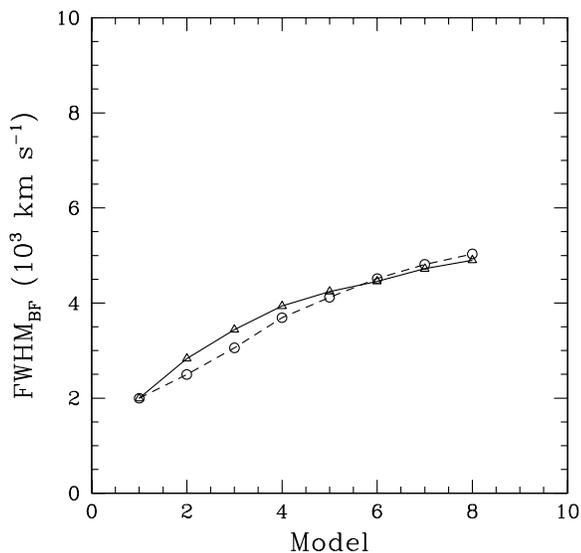}
\caption{Expected FWHM in the BF of emission lines originated in a disk
  structure. The empty triangles connected by the continuous line are the
  predicted values for reference disk models with $i = 10$° in the
  gravitational field of black holes with M = 1, 2, 4, 6, 8, 10, 12,
  and 14 $\cdot 10^7 {\rm M}_\odot$ (respectively models from 1 to 8);
  the circles with the dashed line show the predicted behavior for the
  reference disk model in the gravitational field of the black hole
  with M = $10^7 {\rm M}_\odot$, when seen under inclinations of $i =
  10$°, 15°, 20°, 25°, 30°, 35°, 40°, and 45°
  (again from model 1 to 8). \label{f04}}
\end{figure}
\section{Results and discussion}
A clear determination of the black hole mass is not possible unless we
are able to discriminate the role played by $f$ in Eq.~(1). In many
circumstances, the black hole mass problem is dealt with through the
assumption of a particular geometry, such as a random distribution of
virial motions \citep{Netzer90, Wandelea99, Peterson00}, or a
flattened rotating system with an inclination mostly inferred by means
of statistical considerations \citep{Decarli08a}. Many authors
\citep[etc.]{Vestergaard00, Nikolajuk05, Peterson04, Sulentic06}
pointed out that a considerable flattening is likely to be an
intrinsic property of the BLR structure and even that a simple
assumption about its inclination may not remove the problem. The very
nature of the broad line emitting entities has been investigated
extensively \citep[see, for example,][]{Arav97, Arav98, Laor06},
leading to the conclusion that the broad line profiles either result
from the combination of a large number of emitters, in the order of
$\sim10^7$, or it is produced by motions of a smooth medium. In both
cases, it is noticed that a random motion pattern could not be
dynamically stable.

With the exception of those objects whose line profiles clearly show
double peaks, a strong clue towards an highly inclined rotating
system, the BLR inclination is still an open question of crucial
importance for determination of black hole mass and accretion
rate. Indeed, the assumption of a universal geometrical factor usually 
leads to the detection of high accretion rates in Narrow Line Seyfert
1 galaxies (NLS1) \citep[e. g.][]{Boller96}, while adjusting the
geometrical factor, according to statistics, affects the black hole
masses largely reducing most of the differences. Both these paths
might be sources of systematic misunderstandings, therefore a direct
and independent measurement of the BLR inclination, or, alternatively,
of the black hole accretion rate, would be highly desirable in order to
discriminate among the actual dynamical properties and the effect of
inclination \citep{Kelly08}.

A similar test has been performed, for a restricted sample, by
\citet{Hicks08} in near infrared spectroscopic observations, leading to
the conclusion that the observed gas kinematics is consistent with RM
based results for nearly face-on disk structures.

\subsection{Inclination and line profile broadening}
Since the BLR structure cannot be represented by a random motion
pattern, the shape of the broad emission lines exhibits large
deviations from the Gaussian profile. In the extreme case of the
marked double peaks in the spectral lines of Arp 102B, studied by
\citet{Chen89}, the BLR structure is well explained in terms of the
combination of a quasi spherical component with a rotating disk,
probably the external accretion disk, seen at an inclination of $i =
32$°. \citet{Popovic04} applied the same model to other single peaked
line emitting sources and they found that mildly inclined disks could
be responsible for the observed line profiles as well, although
uncertainties on the model free parameters might affect the values of
the inferred inclinations, as \citet{Collin06} pointed out.

If the BLR has a flattened component which is seen at low inclination, its
emission lines clearly do not exhibit double peaks, but the geometrical
structure still modifies the dynamical interpretation of data. We
illustrate this concept in Fig.~4, where we plot the expected FWHM in
the broadening function of disks surrounding black holes of increasing
mass and we compare it to the situation of a black hole of fixed mass,
but with the disk seen under different inclinations. It is clear that,
within the range of our calculations, there is a mass - inclination
degeneracy on the resulting FWHM.

Exploiting the model developed in \citet{Chenea89}, \citet{Chen89}, and
\citet{Popovic04}, we computed a range of expected non-Gaussian profiles
in the case of a two component BLR structure seen at different
inclinations, with a flattened rotating disk and a surrounding
distribution of gas, giving rise to a bell-shaped contribution. In its
original purpose, this model was conceived to fit the properties of an
accretion disk, introducing some free parameters for the disk radii,
intrinsic velocity dispersion, and line emission. Adjusting these
parameters, it would be possible to fit the broad emission line
profiles of possibly all the spectra of our sample, but reasonable
fits can be obtained in several ways, without tightly constraining the
physical properties of the BLR. Here we try to predict the effect of a
flattened BLR component on the observed line profiles, therefore we
fix some of these parameters on the basis of the results collected by
\citet{LaMura07}. Our reference model assumes $R_{in} = 1834 R_S$ for
inner radius, $R_{BLR} = 18340 R_S$ for outer radius, $\sigma_{Disk} =
0.003 c$ as the intrinsic velocity dispersion in the disk, and $\alpha
= -2.0$ for the radial emission power law, where $R_S$ and $c$
represent the Schwarzshild radius and the speed of light. Moreover,
this model includes a bell-shaped gas distribution having a velocity
dispersion of $\sigma_{Bell} = 0.008 c$. The reference model carries
out the best match to the observed line profiles with the assumption
of various disk inclinations.

\begin{figure}[t]
\includegraphics[width = 8.3cm]{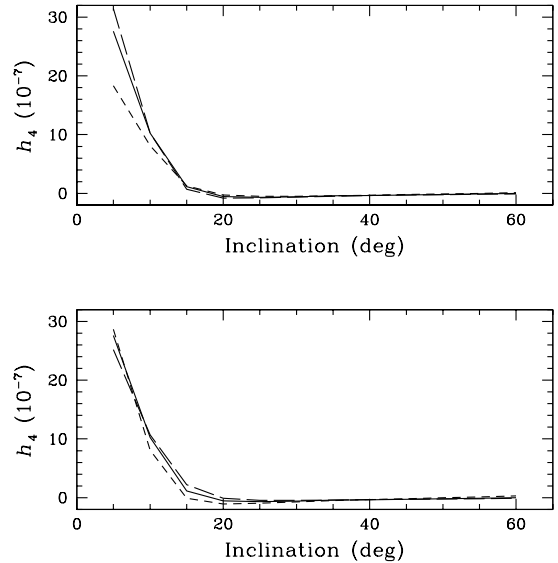}
\caption{Comparison of the reference model predictions concerning the
  broadening function kurtosis (continuous lines) with four variants:
  in the upper panel we plot models with stronger (long dashed line) and
  weaker (short dashed line) disk emission with respect to the bell-shaped
  component; in the bottom panel we show the differences obtained by setting
  $\sigma_{Bell} = 0.07 c$ (short dashed line) and $\sigma_{Bell} = 0.09 c$
  (long dashed line). \label{f05}}
\end{figure}
In Fig.~5, we compare the reference model with some variants, obtained
with slightly different parameters. We note that all the models
predict a strong dependence of the line profile kurtosis (the
coefficient $h_4$ in the Gauss-Hermite expansion) on the disk
inclination, in the range of small values of $i$. The reason is quite
simple, because a nearly face-on disk enhances the low radial velocity
peak of the BF, increasing the kurtosis of the profile, while an
edge-on disk is more likely to affect the high velocity wings.
However, differences in the relative normalization of the bell-shaped
component, with respect to the disk, or in its intrinsic velocity
dispersion, may also affect the inferred kurtosis. The assumed
strength of the bell-shaped component has a large effect on the
predicted kurtosis for $i \leq 10$°, while its velocity dispersion has
a weaker influence in the range 10° $\leq i \leq 20$°. Since large
changes in the model parameters quickly result in predictions that do
not match the observed line profiles, we assume, in our calculations,
a confidential uncertainty of $\Delta i = \pm 2$°, for $i \leq 10$°,
and of $\Delta i = \pm 5$°, for $i > 10$°, where the dependence of
kurtosis on inclination becomes shallower. At $i \geq 20$° the
kurtosis is no longer a useful indicator of inclination.

As we show in Fig.~6, however, where we plot the measured values of
$h_4$ as a function of FWHM$_{\rm H\beta}$, there is a remarkable
evolution of the line profile kurtosis, which decreases for increasing
line profile width. Such an effect is a clear indication that a
considerable variation of the geometrical factor $f$ might be present
and it should be taken into account in order to estimate the actual
properties of the SMBH located in the centre. Using the model
predictions, we can exploit the broad line kurtosis to estimate the
inclination of the flattened BLR component and to apply a correction
to our dynamical interpretation of the observed line profiles. It
should be noted, however, that, although the kurtosis is estimated
from the whole profile, it reduces the available information to a
single parameter. It is, therefore, very important that the model
provides a good fit of the observed line profiles.
\begin{figure}[t]
\includegraphics[width = 8.3cm]{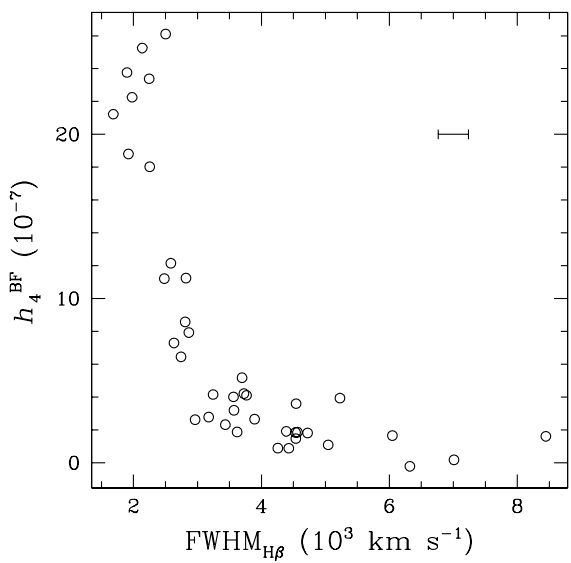}
\caption{BF kurtosis plotted as a function of FWHM$_{\rm H\beta}$. We observe
  a large evolution of the line profile kurtosis as a function of
  broadening, suggesting the possibility of inclination effects in
  nearly face-on flattened structures. Such effects become weaker as
  the profile width increases.\label{f06}}
\end{figure}

\subsection{Mass and accretion rate estimates}
It can be shown that completely neglecting the role played by the BLR
geometrical factor may lead to incorrect black hole mass estimates, with
uncertainties that, in the worst cases, could span over two orders of
magnitude. This problem is particularly important in the case of NLS1
galaxies, whose nature has been carefully investigated, to find out
whether they are characterized by flattened rotating structures seen
at low inclination \citep{Osterbrock85}, or they are actually low mass
black holes accreting at very high rates, sometimes well beyond the
Eddington limit, as it is discussed, for instance, in \citet{Boller96}
or in \citet{Komossa08}. Moreover, the role played by non
gravitational forces, especially in the case of high radiative
efficiency, may also influence the kinematical properties of gas,
as suggested by \citet{Marconi08}, affecting the reliability of the
virial assumption. In their work, \citet{LaMura07} found that, although
NLS1 had quite high accretion rates, they were not exceptional with
respect to other AGN in the sample, a result echoed by the
considerations of \citet{Decarli08a}.

\begin{figure}[t]
\includegraphics[width = 8.3cm]{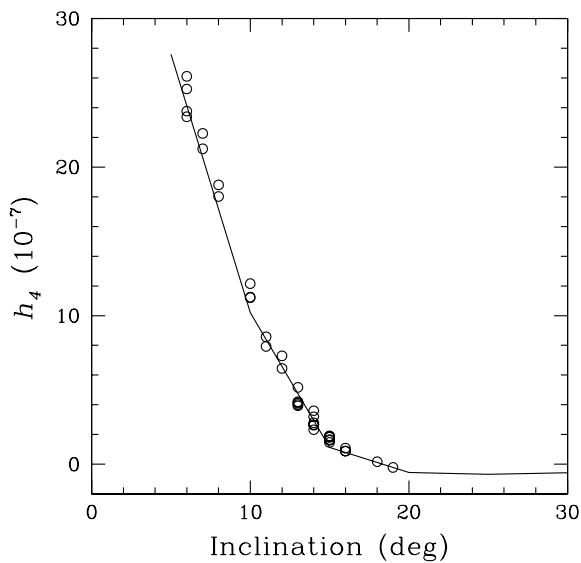}
\caption{BLR inclination inferred by comparison among the BF kurtosis,
  predicted by the reference model for various inclinations, and the
  corresponding distribution observed in our data. According to the
  model predictions, the BF kurtosis is very sensitive to flattening
  and inclination in nearly face-on structures, while it becomes a
  weaker indicator for larger inclinations. \label{f07}}
\end{figure}
On the other hand, while \citet{Shemmer06} argue that X-ray observations may
provide a direct clue to the black hole accretion rate and thus remove the
degeneracy introduced by the FWHM$_{\rm H\beta}$ dependent mass estimates,
\citet{Decarli08b} and \citet{Labita06} use the black hole correlations with
the host properties, identified by \citet{Ferrarese00, Ferrarese06},
to calibrate the geometrical factor. Both methods suggest that some
care should be taken in using only the profile of H$\beta$ to infer
the physical properties of AGN.

With the information coming from the line profile distortions, we compute our
estimates of the black hole mass and accretion rate by introducing an {\it
  equivalent velocity field}, defined by:
$$v_{eq} = \frac{1}{2}\left[\frac{\sqrt{3}}{2} {\rm FWHM}_{Bell}({\rm H}\beta)
  + \frac{{\rm FWHM}_{Disk}({\rm H}\beta)}{4\sin i}\right]. \eqno(19)$$
Assuming that the line profile broadening results from both planar and
non-planar motions \citep[][etc.]{Labita06, McLure02, Jarvis06},
$v_{eq}$ combines the velocity estimates obtained from the H$\beta$
emission line profile by fitting two Gaussian functions, which are
subsequently compared with the reference model, providing a
distinction among the bell-shaped and the flattened contributions. The
corresponding geometrical factors are assumed to be, respectively, the
classical interpretation of \citet{Netzer90} and that of a
rotating disk, confined in a smaller region with respect to the other
component. The inclination of the disk is estimated by comparison of
the BF kurtosis with that of the reference model, as shown in Fig.~7,
and its characteristic radius is assumed to be approximately four
times smaller than the typical size of the bell-shaped component. To
calculate the black hole mass, we introduce $v_{eq}$ in Eq.~(1),
bringing the geometrical factor into the modified velocity field, and
we estimate the corresponding Eddington ratio from the bolometric
luminosity in Eq.~(18).

The results of our measurements and calculations are summarized in
Table~2, together with the computed uncertainty ranges.  Our method
breaks the strong dependence of the black hole accretion rate and mass
estimates on the width of H$\beta$ since it exploits more indications,
coming from the broadening functions of the observed Balmer
lines. Moreover, the typical inclinations inferred for the BLR of our
spectra are consistent with those estimated by \citet{Popovic08},
suggesting that this situation is quite common in single peaked broad
line emitters. Table~2, however, does not include the errors which
could be introduced by our assumptions, concerning the source
luminosity and the BLR structure. Such uncertainties may be as large
as a factor of 2 or 3, as seen in the AGN SED distribution, or by
adopting different structural models for the BLR. We shall further
discuss the role of these uncertainties with the help of some
consistency checks.

\begin{deluxetable*}{lcccccc}
\tabletypesize{\footnotesize}
\tablewidth{0pt}
\tablecaption{Estimated properties of the sources \label{tab02}}
\tablehead{\colhead{Name\tablenotemark{a}} & \colhead{L$_{bol}$} &
  \colhead{$h_3$} & \colhead{$h_4$} & \colhead{$i$} &
  \colhead{M$_{\rm BH}$} & \colhead{L$_{bol}$ /
    L$_{Edd}$}\\ \colhead{} & \colhead{($10^{44} {\rm erg s}^{-1}$)} &
  \colhead{($10^{-7}$)}  & \colhead{($10^{-7}$)} & \colhead{(deg)} &
  \colhead{($10^8 {\rm M_\odot}$)} &  \colhead{}}
\startdata
RX J0801.5+4736\polstar          &  20 $\pm$ 1 &   3.1 &  0.2 & 18 & 14 $\pm$  5 & 0.011 $\pm$ 0.004 \\
RXS J080358.9+433248             &  36 $\pm$ 2 &   8.9 &  7.3 & 12 &  5 $\pm$  2 & 0.054 $\pm$ 0.027 \\
1RXS J080534.6+543132            &  83 $\pm$ 2 & -71.6 &  2.8 & 14 &  8 $\pm$  3 & 0.083 $\pm$ 0.032 \\
SDSS J081222.99+461529.1         &  37 $\pm$ 1 &  68.6 &  8.6 & 11 & 10 $\pm$  4 & 0.028 $\pm$ 0.013 \\
2MASSi J0816522+425829           &  25 $\pm$ 1 &  -0.3 &  4.1 & 13 &  9 $\pm$  4 & 0.021 $\pm$ 0.010 \\
NGC 2639 U10                     &  17 $\pm$ 1 &  46.8 &  1.7 & 15 &  4 $\pm$  1 & 0.031 $\pm$ 0.012 \\
SDSS J085632.39+504114.0         &  39 $\pm$ 1 &  39.3 & 11.2 & 10 &  5 $\pm$  3 & 0.057 $\pm$ 0.036 \\
SDSS J085828.69+342343.8         &  37 $\pm$ 1 & -15.4 &  3.2 & 14 &  7 $\pm$  3 & 0.038 $\pm$ 0.016 \\
SDSS J090455.00+511444.6         &  19 $\pm$ 1 &  45.0 &  1.9 & 15 &  7 $\pm$  3 & 0.021 $\pm$ 0.009 \\
RX J0906.0+4851                  &  35 $\pm$ 1 & 197.8 & 26.1 &  6 & 13 $\pm$  7 & 0.020 $\pm$ 0.011 \\
RX J0908.7+4939                  &  87 $\pm$ 2 &  83.1 & 25.3 &  6 & 14 $\pm$  6 & 0.048 $\pm$ 0.023 \\
1WGA J0931.9+5533\polstar        &  26 $\pm$ 1 &  -4.7 &  1.1 & 16 & 32 $\pm$  9 & 0.006 $\pm$ 0.002 \\
SDSS J093653.84+533126.8         &  25 $\pm$ 1 & -54.9 &  0.9 & 16 & 36 $\pm$ 12 & 0.005 $\pm$ 0.002 \\
FIRST J094610.9+322325\polstar   &  31 $\pm$ 2 &-161.1 & 21.2 &  7 &  5 $\pm$  2 & 0.048 $\pm$ 0.021 \\
KUV 09484+3557                   &  22 $\pm$ 1 &  49.4 &  4.1 & 13 &  5 $\pm$  2 & 0.037 $\pm$ 0.017 \\
HS 1001+4840                     &  24 $\pm$ 3 &  58.8 &  2.6 & 14 & 12 $\pm$  4 & 0.016 $\pm$ 0.008 \\
PC 1014+4717                     &  25 $\pm$ 1 &  85.5 & 12.1 & 10 &  5 $\pm$  3 & 0.038 $\pm$ 0.023 \\
RX J1030.4+5516\polstar          &  40 $\pm$ 4 & 144.8 & 23.4 &  6 & 12 $\pm$  6 & 0.026 $\pm$ 0.016 \\
FBQS J103359.4+355509\polstar    &  13 $\pm$ 1 &  26.7 &  3.6 & 14 &  3 $\pm$  1 & 0.030 $\pm$ 0.010 \\
SBS 1047+557B                    &  43 $\pm$ 2 &  44.6 & 18.0 &  8 &  8 $\pm$  3 & 0.039 $\pm$ 0.016 \\
RX J1054.7+4831\polstar          &  49 $\pm$ 2 &  31.1 &  0.9 & 16 & 12 $\pm$  5 & 0.030 $\pm$ 0.013 \\
FBQS J105648.1+370450\polstar    &  20 $\pm$ 2 &  19.8 &  1.5 & 15 &  5 $\pm$  2 & 0.028 $\pm$ 0.014 \\
FBQS J110704.5+320630\polstar    &  18 $\pm$ 1 & -35.5 &  1.6 & 15 & 24 $\pm$ 11 & 0.006 $\pm$ 0.003 \\
FBQS J112956.5+364919\polstar    &  10 $\pm$ 1 &  67.7 &  4.0 & 13 &1.8 $\pm$ 0.7& 0.041 $\pm$ 0.020 \\
FBQS J115117.7+382221\polstar    & 188 $\pm$ 4 & -17.8 &  5.2 & 13 & 22 $\pm$  8 & 0.065 $\pm$ 0.026 \\
RX J1200.4+3334                  &   9 $\pm$ 1 &  10.8 &  7.9 & 11 &  3 $\pm$  1 & 0.023 $\pm$ 0.014 \\
RX J1203.8+3711\polstar          &  27 $\pm$ 2 &  79.2 &  4.2 & 13 &  3 $\pm$  2 & 0.068 $\pm$ 0.041 \\
1RXS J121759.9+303306            &  17 $\pm$ 1 &  33.4 &  2.3 & 14 &  5 $\pm$  2 & 0.025 $\pm$ 0.013 \\
RX J1218.3+3850                  &  15 $\pm$ 1 & -17.4 &  1.9 & 15 &  6 $\pm$  3 & 0.018 $\pm$ 0.009 \\
FBQS J122035.1+385317\polstar    &  93 $\pm$ 2 & -84.3 & 23.8 &  6 &  9 $\pm$  4 & 0.079 $\pm$ 0.040 \\
FBQS J122424.2+401510\polstar    &  35 $\pm$ 2 &  15.0 &  1.8 & 15 &  8 $\pm$  2 & 0.035 $\pm$ 0.013 \\
FBQS J122624.2+324429\polstar    &   7 $\pm$ 1 &  12.7 &  1.8 & 15 &  3 $\pm$  1 & 0.016 $\pm$ 0.007 \\
FBQS J125602.0+385230            &  31 $\pm$ 3 & -78.5 & 22.3 &  7 &  4 $\pm$  2 & 0.054 $\pm$ 0.032 \\
FBQS J132515.0+330556            &  27 $\pm$ 1 & -29.5 & 18.8 &  8 &  5 $\pm$  2 & 0.039 $\pm$ 0.017 \\
SDSS J144050.77+520446.0         &  20 $\pm$ 1 &  67.7 & 11.2 & 10 &  6 $\pm$  3 & 0.028 $\pm$ 0.016 \\
RX J1452.4+4522\polstar          &  88 $\pm$ 4 &   4.8 & -0.2 & 19 &  6 $\pm$  2 & 0.105 $\pm$ 0.044 \\
FBQS J145958.4+333701\polstar    &  89 $\pm$ 20&  41.7 &  1.9 & 15 &  8 $\pm$  4 & 0.081 $\pm$ 0.056 \\
FIRST J154348.6+401324\polstar   &  29 $\pm$ 2 & -53.0 &  3.9 & 13 &  5 $\pm$  2 & 0.042 $\pm$ 0.016 \\
SDSS J154833.03+442226.0\polstar &  23 $\pm$ 1 &  -6.6 &  2.6 & 14 &  8 $\pm$  3 & 0.021 $\pm$ 0.009 \\
FBQS J155147.4+330007\polstar    &  32 $\pm$ 3 &  12.0 &  6.4 & 12 &  3 $\pm$  2 & 0.070 $\pm$ 0.039 \\
\enddata
\tablenotetext{a}{Objects marked with a \polstar\ have a polarization
  measurement in the NVSS catalogue.}
\tablecomments{This table reports the bolometric luminosities, the
  Gauss-Hermite expansion coefficients for the BF profile, the
  inferred inclination, and an estimate of the central black hole mass
  and accretion rate.}
\end{deluxetable*}
\subsection{Discussion}
While our estimates of bolometric luminosity and, consequently, black
hole mass and accretion rate are essentially scaled by our measurement of the
optical continuum luminosity, the dynamical interpretation of the line
profiles still suffers from undeniable shortcomings. Adopting a two component
model to explain the line profile broadening complicates the relationship
among FWHM$_{H\beta}$ and the black hole mass, introducing a geometrical
factor which depends on the inclination of the flattened component and on its
relative importance with respect to the BLR as a whole. Because disks
are the most viable solution to support accretion flows in presence of
angular momentum, numerous authors suggested that the broad line gas
could originate in the disks themselves \citep[e. g.][]{Shields77,
  Shlosman85, Emmering92}. Models based on accretion disks only,
however, have great difficulties in accounting for AGN observational
properties \citep[see][for example]{Kinney94}. The assumption of a two
component model improves our ability to understand the observed line
profiles, but it still fails in placing strong constraints on the
structure of the BLR, since the origin of the bell-shaped component is
not clear. Indeed, there are models, such as those of
\citet{Collin-Souffrin90}, \citet{Jackson91}, or \citet{Murray97},
which achieve a good match with observations on more physical grounds,
either exploiting very large disk radii, or computing the effect of
radiation transfer across radial gas flows close to the disk. Clearly,
the choice of different models affects the interpretation of AGN
dynamical properties and this is a major concern in the case of the
BLR.

A particularly important problem, involving the determination of AGN
physical properties from emission lines, resides in the line profile
asymmetries. Several factors, such as partial obscuration,
geometrical structure, or large scale non-virialized motions can
produce asymmetric line profiles. Moreover, relativistic effects
within the gravitational field of the SMBH give raise to asymmetries,
especially in the high velocity wings of the profile, which are
included in the calculations of the model by \citet{Chen89}. In order
to assess how much the asymmetric component affects our estimates of
the velocity field, we introduced an asymmetry parameter:
$$K_3 = h_3 H_3({\rm HWHM_{H\beta}}), \eqno(20)$$
expressing the relative contribution of the asymmetric component, with
respect to the Gaussian component, in the profile of H$\beta$ at its
half-maximum level. As we show in Fig.~8, the asymmetric component
gives a contribution to the FWHM which rarely exceeds the 10\%
level. The most extreme cases, where the asymmetric component becomes
larger than 20\%, occur only in the range of very broad line emitting
sources. Although this does not appear to be a general property of
broad line objects, it echoes the observation of larger asymmetries in
objects where ${\rm FWHM_{H\beta}} > 4000\, {\rm km\, s^{-1}}$, which
is among the features identified by \citet{Sulentic00, Sulentic06} in
their distinction between Population A and B sources. Objects with the
largest asymmetries are more problematic for the comparison with the
reference model, used to calculate the equivalent velocity field in the
BLR. However, comparing their masses and accretion rates with
those of the other sources, we do not find systematic differences that
could suggest the need for specific model corrections in the asymmetric
line emitters, as it is shown in Fig.~9. Here we see that the result
of introducing the BLR inclination in our calculations is to remove the
strong dependence of the accretion rate on FWHM$_{\rm   H\beta}$, that was
commonly found with isotropic mass estimates. Instead, we are left with
a much more complex situation, where, though a slight trend to measure
lower accretion rates in broad line emitting sources is still present,
it is not a universal condition. An inverse power law fit to the data
yields $L_{bol} / L_{Edd} \propto {\rm   FWHM_{H\beta}}^{-0.75}$,
considerably weaker than the old isotropic prediction of $L_{bol} /
L_{Edd} \propto {\rm FWHM_{H\beta}}^{-2}$. In particular, we do not
observe dramatic excesses in the accretion rate of our sources, whose
estimates are far below the corresponding Eddington limits.

\begin{figure}
\includegraphics[width = 8.3cm]{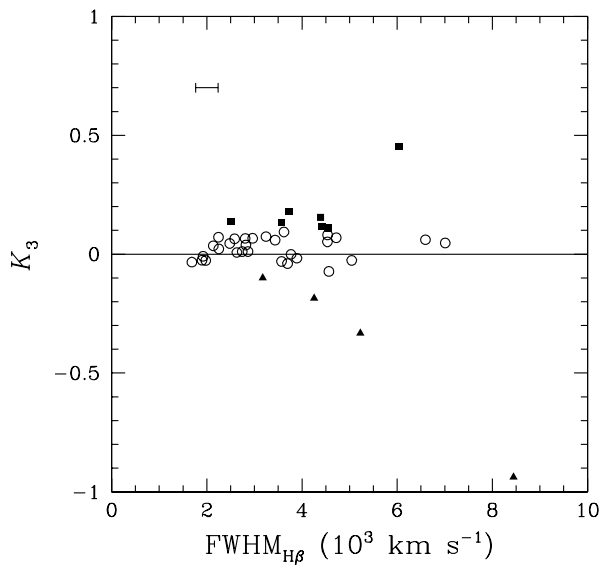}
\caption{H$\beta$ asymmetry parameter distribution with respect to
  FWHM$_{\rm H\beta}$. Filled symbols represent objects where the
  asymmetric component exceeds 10\% of the Gaussian contribution.
  We plot as triangles objects that are affected by negative
  asymmetry, yielding blue shifted peaks and red shifted wings, and
  squares for positive asymmetry sources, having a red shifted peak
  with blue shifted wings. Large asymmetries characterize objects
  where fits with the reference model are more likely to be
  problematic. The bar in the upper left region of the diagram is a
  median estimate of the measurement errors. \label{f08}}
\end{figure}
Most of the results achieved in this work depend critically on the
choice of our reference model, which leads us to conclude that the
BLR has a flattened component, that is commonly seen at $i \leq
20$°. In the case of radio-loud sources, nearly face-on disk
structures are likely to produce a radio jet oriented along our line
of sight towards the object, and the resulting signal should be highly
variable and polarized. Although we were not able to find any
information about variability, some of the radio loud sources in our
sample have been detected in the NRAO VLA Sky Survey (NVSS), which
provides measurments of the radio flux and polarization at the
frequency of 1.4~GHz \citep{Condon98}.\footnotemark
\footnotetext{Polarization data are available at
  http://www.cv.nrao.edu/nvss/NVSSlist.shtml} We identify these
objects in Table~2 and we compare the degree of linear polarization
with our inclination estimates in Fig.~10. Although the uncertainties
are quite large, a significant degree of linear polarization is
detected in many objects and it appears to be an averagely decreasing
function of $i$.

A comparison of our mass determinations with the old isotropic
assumption allows us to study the properties of the geometrical factor
within our sample. The situation depicted in Fig.~11 clearly indicates
that significant effects, up to a factor $\sim 30$, should be expected
and that they are more commonly observed in the range of sources with
FWHM$_{\rm H\beta} \leq 3000 - 4000 {\rm km\, s^{-1}}$. We find that the
average value of the geometrical factor for black hole mass
determinations based on FWHM$_{\rm H\beta}$ is $f = 10.58 \pm 7.70$,
marginally consistent with the result achieved by \citet{Onken04}, who
gave $f = 5.5 \pm 1.9$ using the emission line dispersions.

\section{Conclusions}
In this work we investigated the shape of the emission line broadening
function in the BLR of active galactic nuclei. We used a technique
based on cross-correlation and Gauss-Hermite line profile fitting,
applied to the Balmer series, to infer the broadening functions and we
compared them with the predictions of a structural model for the
BLR. According to our results, we come to the following conclusions:
\begin{figure}[t]
\includegraphics[width = 8.3cm]{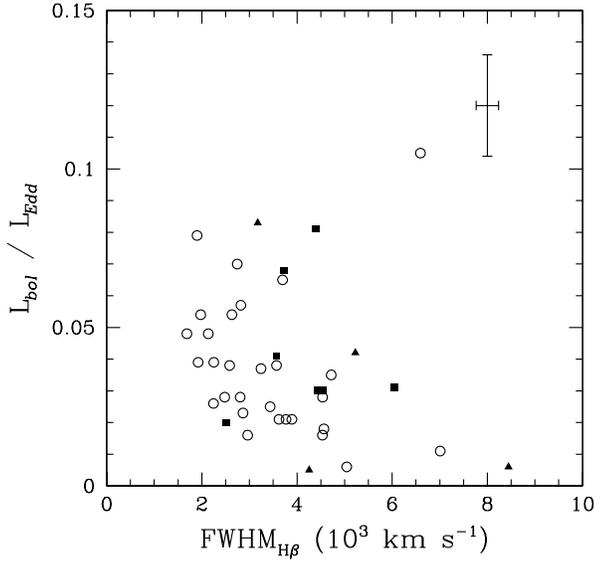}
\caption{SMBH accretion rates as a function of FWHM$_{\rm H\beta}$, with the
  same symbols as in Fig.~8. The cross in the upper right region of
  the diagram is the median uncertainty of measurements. We do not
  detect either dramatic accretion rate excesses or systematic trends
  associated to the line width, although higher accretion rates appear
  to be more common in the domain of narrow emission line
  sources. This conclusion seems not to be affected by asymmetric line
  profiles. \label{f09}}
\end{figure}
\begin{itemize}
\item the line profile broadening functions carry much detailed information
  about kinematics of the BLR, which can be better understood by means of
  techniques exploiting the whole profile, rather than restricting on specific
  parameters;
\item the observed distribution of line profile kurtosis is consistent
  with the presence of a flattened component in the BLR, with a
  typical inclination $i \leq 20$°, though the actual values of $i$
  may depend on the adopted model;
\item some of the objects included in the sample have quoted
  measurments of linear polarization at radio frequencies, which
  averagely increase as the estimated BLR inclination approaches face-on;
\item correcting the SMBH mass and accretion rate estimates for geometry
  reduces the anti-correlation among FWHM$_{\rm H\beta}$ and accretion
  rate to a much weaker trend;
\item there are no particular indications, in our results, for a
  strong influence of the line profile asymmetries on the determination
  of the SMBH properties.
\end{itemize}

Although this analysis may represent an advance in the problem of determining
the role of the BLR geometrical factor, more questions should be answered,
concerning how the two components combine in the observed line profiles. A
precious contribution in this effort would probably result from the
comparison of this technique with some independent way to estimate the black
hole accretion rate. Recent works suggested that this test could be possible,
for example, with X-ray observations. The ability to constrain the BLR
geometrical factor, then, could be applied to study the properties of black
hole - host galaxy scaling relations with improved accuracy.
\begin{figure}[t]
\includegraphics[width = 8.3cm]{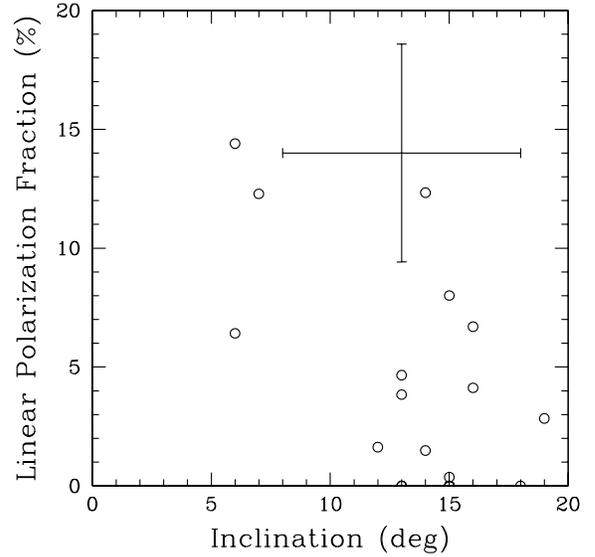}
\caption{Degree of linear polarization at the radio frequency of
  1.4~GHz as a function of the inferred BLR inclination. The cross in
  the upper left corner gives the median uncertainty
  estimate. Polarization data are from the NVSS catalogue.\label{f10}}
\end{figure}
\begin{figure}[t]
\includegraphics[width = 8.3cm]{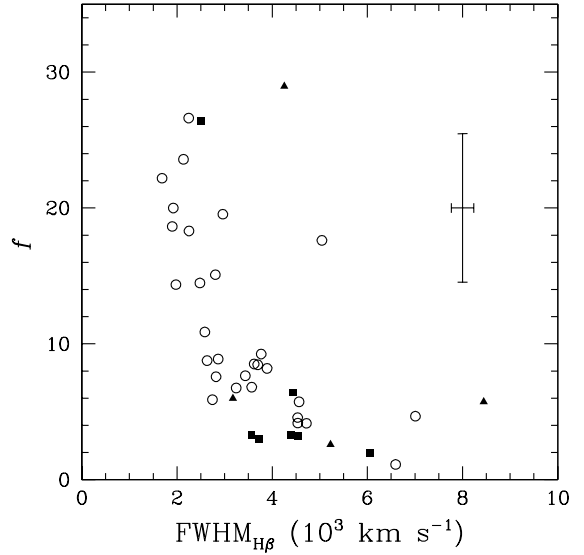}
\caption{Geometrical factors obtained by comparison of the virial
  product of Eq.~(1) and our black hole mass estimates, with the same
  symbols as in Fig.~8. The effect of the geometrical factor is
  generally stronger as the line profile width decreases, with a
  correction to the black hole mass estimate which can be as high as a
  factor of $\sim 30$.\label{f11}}
\end{figure}

\vspace{0.5cm}
\noindent We thank the anonymous referee for useful suggestions and
discussion leading to significant improvement of this work.

L. \v C. Popovi\'c was supported by the Ministry of Science
of R. Serbia through project 146002 ``Astrophysical Spectroscopy of
Extragalactic Objects.''

Funding for the SDSS and SDSS-II has been provided by the Alfred
P. Sloan Foundation, the Participating Institutions, \vspace{10cm}
the National Science Foundation, the U.S. Department of Energy, the
National Aeronautics and Space Administration, the Japanese
Monbukagakusho, the Max Planck Society, and the Higher Education
Funding Council for England. The SDSS Web Site is
http://www.sdss.org/.

The SDSS is managed by the Astrophysical Research Consortium for
the Participating Institutions. The Participating Institutions are
the American Museum of Natural History, Astrophysical Institute
Potsdam, University of Basel, University of Cambridge, Case
Western Reserve University, University of Chicago, Drexel
University, Fermilab, the Institute for Advanced Study, the Japan
Participation Group, Johns Hopkins University, the Joint Institute
for Nuclear Astrophysics, the Kavli Institute for Particle
Astrophysics and Cosmology, the Korean Scientist Group, the
Chinese Academy of Sciences (LAMOST), Los Alamos National
Laboratory, the Max-Planck-Institute for Astronomy (MPIA), the
Max-Planck-Institute for Astrophysics (MPA), New Mexico State
University, Ohio State University, University of Pittsburgh,
University of Portsmouth, Princeton University, the United States
Naval Observatory, and the University of Washington.

\end{document}